\documentclass[myepj-spec,pdftex]{mySvjour}
\usepackage{graphicx}
\usepackage{hyperref}
\usepackage{color}
\usepackage{xspace}
\usepackage[square,sort&compress,numbers]{natbib}   
\usepackage{verbatim}
\usepackage{amsmath,amssymb,amsfonts} % Typical maths resource packages
\usepackage{tabularx}

\usepackage{lmodern,bm}                
\usepackage[T1]{sansmath} 
\SetMathAlphabet{\mathsfbf}{sans}{\sansmathencoding}{\sfdefault}{bx}{sl}
\usepackage{etoolbox}
\AtBeginEnvironment{sansmath}{}{}{}

\usepackage{lipsum}

\newcommand\blfootnote[1]{%
  \begingroup
  \renewcommand\thefootnote{}\footnote{#1}%
  \addtocounter{footnote}{-1}%
  \endgroup
}

\definecolor{darkblue1}{rgb}{0,0,.2}
\definecolor{darkblue}{rgb}{0,0,.2}
\definecolor{darkred}{rgb}{0.5,0,0}
\pagecolor{white} % Background color
\color{black}     % Text color
\hypersetup{breaklinks=true, 
            colorlinks=true, 
            linkcolor=darkblue1, 
            menucolor=darkblue1, 
            urlcolor=darkblue1,
            citecolor=darkblue1,
            pdftitle={},
            pdfauthor={},
            pdfsubject={},
            pdfkeywords={},
            pdfproducer={}
}
\bibstyle{plain}
\newcommand{\bi}{\begin{itemize}}
\newcommand{\ei}{\end{itemize}}

\newcommand{\ben}{\begin{enumerate}}
\newcommand{\een}{\end{enumerate}} 

\newcommand{\bt}[1]{\begin{table}[tb]\begin{tabular}{#1} \hline\hline  \\[-1.0em]}
\newcommand{\et}[2]{\hline\hline \end{tabular} \caption{#1} \label{#2} \end{table}}

\newcommand{\be}{\begin{equation}}
\newcommand{\ee}{\end{equation}}

\newcommand{\bea}{\begin{eqnarray}}
\newcommand{\eea}{\end{eqnarray}}

\newcommand{\bc}{\begin{comment}}
\newcommand{\ec}{\end{comment}}

\newcommand{\chisq}{\ensuremath{\chi^2/\text{n.d.f.}}\xspace}
\newcommand{\chisqsmall}{\ensuremath{\chi^2/\text{ndf}}\xspace}
\newcommand{\chiS}{\ensuremath{\chi^2}\xspace}

\newcommand{\epemHadr}{\ensuremath{e^+e^-\to{\rm Hadrons}}\xspace}
\newcommand{\pp}{\ensuremath{\pi^+\pi^-}\xspace}
\newcommand{\ppy}{\ensuremath{\pi^+\pi^-(\gamma)}\xspace}

\newcommand{\mev}{\ensuremath{\mathrm{\,Me\kern -0.1em V}}\xspace}
\newcommand{\gev}{\ensuremath{\mathrm{\,Ge\kern -0.1em V}}\xspace}

\newcommand{\Qs}{\ensuremath{{\rm Q}^2}\xspace}

\newcommand{\mW}{\ensuremath{M_{W}}\xspace}
\newcommand{\MW}{\ensuremath{M_{W}}\xspace}
\newcommand{\mZ}{\ensuremath{M_{Z}}\xspace}
\newcommand{\MZ}{\ensuremath{M_{Z}}\xspace}
\newcommand{\mH}{\ensuremath{M_{H}}\xspace}
\newcommand{\MH}{\ensuremath{M_{H}}\xspace}

\newcommand{\mt}{\ensuremath{{m}_t}\xspace}

\newcommand{\amu}{\ensuremath{a_\mu}\xspace}

\newcommand{\amuSM}{\ensuremath{a_\mu^\text{SM}}\xspace}
\newcommand{\amuSMpheno}{\ensuremath{a_\mu^\text{SM (Pheno)}}\xspace}

\newcommand{\amuexp}{\ensuremath{a_\mu^\text{Exp}}\xspace}

\newcommand{\amuHVPLO}{\ensuremath{a_\mu^\text{HVP, LO}}\xspace}

\newcommand{\amuHVPLOpheno}{\ensuremath{a_{\mu\,(\text{Pheno})}^\text{HVP, LO}}\xspace}
\newcommand{\amuHVPLOlatt}{\ensuremath{a_{\mu\,(\text{Lattice})}^\text{HVP, LO}}\xspace}

\newcommand{\alphaQED}{\ensuremath{\alpha_\text{QED}}\xspace}

\newcommand{\DaHadZ}{\ensuremath{\Delta\alpha_{\rm had}(\mZ^2)}\xspace}
\newcommand{\DaHadZpr}{\ensuremath{\Delta^{\prime}\alpha_{\rm had}(\mZ^2)}\xspace}
\newcommand{\DaHadZfive}{\ensuremath{\Delta\alpha^{(5)}_{\rm had}(\mZ^2)}\xspace}

\newcommand{\NP[1]}{\ensuremath{\rm NP_{#1}}\xspace}
\newcommand{\approach[1]}{\ensuremath{\rm \it Approach \, {#1} }\xspace}

\newcommand{\sintheta}{\ensuremath{\sin\!^2\theta_{W}}\xspace}
\newcommand{\sinleff}{\ensuremath{\seffsf{l}}\xspace}
\newcommand{\sinfeff}{\ensuremath{\seffsf{f}}\xspace}
\newcommand{\seffsf}[1]{\sin\!^2\theta^{#1}_{{\rm eff}}}

\newcommand{\aS}{\ensuremath{\alpha_{S}}\xspace}
\newcommand{\aSmZ}{\ensuremath{\alpha_{S}(\mZ^2)}\xspace}

\begin{document}

\twocolumn[{%
  \begin{@twocolumnfalse}

    \begin{flushright}
      \normalsize
%      \today
    \end{flushright}

    \vspace{-2cm}

\title{\Large\boldmath Impact of correlations between \amu and \alphaQED on the EW fit}

\author{Bogdan Malaescu \inst{1}$^{,a}$  \and Matthias Schott \inst{2}$^{,b}$}
\institute{LPNHE, Sorbonne Universit\'e, Universit\'e de Paris, CNRS/IN2P3, Paris, France 
           \and Institute of Physics, Johannes Gutenberg University, Mainz, Germany}

% \noaffiliation

\abstract{
We study the potential impact on the electroweak~(EW) fits due to the tensions between the current determinations of the hadronic vacuum polarisation~(HVP) contributions to the anomalous magnetic moment of the muon~(\amu), based on either phenomenological dispersion integrals using measured hadronic spectra or on Lattice QCD calculations.
The impact of the current tension between the experimental measurement of \amu and the total theoretical prediction based on the phenomenological calculations of the HVP is also studied.
The correlations between the uncertainties of the theoretical predictions of \amu and of the running of $\alpha_{\rm QED}$ are taken into account in the studies.
We conclude that the impact on the EW fit can be large in improbable scenarios involving global shifts of the full HVP contribution, while it is much smaller if the shift is restricted to a lower mass range and/or if the shift in \alphaQED is obtained from that in \amu through appropriate use of the correlations.
Indeed, the latter scenarios only imply at most a $2.6/16$ increase in the $\chisq$ of the EW fits and relatively small changes for the resulting fit parameter values.
}

\maketitle
  \end{@twocolumnfalse}
}]

\section{Introduction}

\blfootnote{$^a$ Bogdan.Malaescu@cern.ch}\blfootnote{$^b$ Matthias.Schott@cern.ch}

A long-standing discrepancy of about 3-4 standard deviations has been observed between the experimental measurement of the anomalous magnetic moment of the muon~(\amuexp)~\cite{Bennett:2006fi} and its Standard Model prediction~(\amuSM) \cite{Aoyama:2020ynm,Davier:2019can,Davier:2017zfy,Keshavarzi:2018mgv,Keshavarzi:2019abf,Colangelo:2018mtw,Hoferichter:2019mqg}.
In this comparison, the leading order hadronic vacuum polarisation part~(\amuHVPLO), derived phenomenologically through dispersion integrals using as input experimental data of \epemHadr\linebreak~(\amuHVPLOpheno), yields the dominant uncertainty of the total theoretical prediction of \amu based on such an approach~(\amuSMpheno).
Recently, the BMW collaboration has achieved an unprecedented sub-percent level precision for a QCD+QED Lattice calculation of this same contribution~(\amuHVPLOlatt)~\cite{Borsanyi:2020mff}.
While yielding a reduced tension between the experimental measurement and the theoretical prediction, this new calculation is in tension with the phenomenological one based on dispersion integrals.
Recent studies indicate that the latter tension seems to originate from the low energy region~(see Ref.~\cite{LaurentSeminar2020} and the updated version of Ref.~\cite{Borsanyi:2020mff}).
At the same time, it has been shown in Ref.~\cite{Campanario:2019mjh} that the last set of NLO radiative corrections for the pion form factor, not considered previously in the event generators used by the experiments, cannot explain the tension between \amuexp and \amuSM.
Comparisons among the \amuHVPLOlatt results obtained by various collaborations, as well as with the \amuHVPLOpheno calculations, have also been performed~\cite{Lehner:2020crt}, using in particular a window method with smoothed steps at the boundaries~\cite{Bernecker:2011gh,Lehner:2017kuc,Blum:2018mom}.
The high precision achieved for the recent result of the BMW collaboration motivates its use as reference \amuHVPLOlatt value in the current study.

It has been advocated in the past that a change of the hadronic spectra~(and hence of~\amuHVPLO) to reduce the tension between the experimental measurement and the theoretical prediction of \amu could introduce tensions in the EW fit~\cite{Passera:2008jk,Crivellin:2020zul}.
More recently, while our work was under completion, it was pointed that a change of the hadronic spectra in the low energy region~(below $0.7\gev$) could allow to reduce the tension for \amu without having too strong an impact on the EW fit~\footnote{The studies in Ref.~\cite{Crivellin:2020zul} also include a scenario where a shift of the hadronic spectra is only applied to the energy region below $1.94\gev$.}, although this would be improbable given the current precision of the data~\cite{Keshavarzi:2020bfy}.
At the same time, in a different study, model-independent bounds were set on the impact that the discrepancy between \amuHVPLOpheno and \amuHVPLOlatt can have on the running of \alphaQED to the Z mass~(\DaHadZ)~\cite{deRafael:2020uif}.

We study these aspects taking into account, to our knowledge for the first time, the full correlations between the uncertainties of the HVP contributions to \amuHVPLOpheno and to \DaHadZ.
Indeed, these correlations are induced by the use in the two dispersion integrals of the same hadronic spectra, perturbative QCD~(pQCD) calculations and narrow resonance contributions.
They have been evaluated in Ref.~\cite{Davier:2019can},~\footnote{Such correlations had also been estimated in Ref.~\cite{Erler:2000nx}, based on the evaluation from Ref.~\cite{Davier:1998si} for the hadronic contribution from the energy threshold up to 1.8$\gev$, completed with the contributions from higher masses evaluated with a sum rule approach.} taking into account in particular the correlations of the (statistical and systematic) uncertainties between the different points/bins of a measurement in a given hadronic channel, between different measurements in the same channel, a well as between different channels.
This evaluation also fully accounts for the tension between the measurements at the BaBar~\cite{Lees:2012cj,Aubert:2009ad} and KLOE~\cite{Ambrosino:2008aa,Ambrosino:2010bv,Babusci:2012rp,Anastasi:2017eio} experiments in the dominant \ppy channel, both through a local re-scaling of the uncertainties by a factor $\sqrt{\chisq}$ and by taking into account the systematic differences between the two measurements~(i.e. comparing the combined dispersion integrals obtained when excluding either BaBar or KLOE).
\section{Description of the EW Fit}

The idea of the global EW fit is to compare the state-of-the-art calculations of EW precision observables with the latest experimental data and thus test the consistency of the Standard Model.
The starting point is the EW sector of the Standard Model that can be described by the masses of the EW gauge bosons $m_V$, the mass of the Higgs boson $\MH$, the EW mixing angle $\theta_W$, as well as the coupling parameters $\alpha_{em} = e^2 / (4 \pi)$ of the electromagnetic interaction, $g$ and $g'$ for the weak interaction as well as the Higgs potential parameter $\lambda$. 
The beauty of the EW theory lies in the predicted relations of its parameters, i.e. the fact that not all of its parameters can be chosen independently from each other. The weak mixing angle, for instance, can be expressed at tree-level as 
\begin{equation}
\label{eqn:sintreelevel}
\sintheta = \left( 1 - \frac{\MW^2}{\MZ^2} \right) ,
\end{equation}
while the mass of the W boson~(\MW) is related to the Fermi constant and the fine-structure constant via

\begin{equation}
\label{eqn:mwtreelevel}
\MW^2 = \frac{\alphaQED \pi}{\sqrt{2} \cdot G_F \cdot (1-\MW^2/\MZ^2)}.
\end{equation}

Hence, at tree level only three free parameters are required.
A common choice of the observables, which are used for the predictions, are those with the smallest experimental uncertainties, i.e. the fine structure constant $\alphaQED$, the $Z$ boson mass \MZ and the Fermi constant $G_F$.
Knowing these, the observables of the EW sector, in particular \MW and \sintheta, can be predicted and confronted with experimental results.
However, just using these tree-level relations will lead to immediate incompatibilities with the respective measurements, since higher order EW corrections have to be taken into account. 
These EW corrections can be formally absorbed into form factors, denoted by $\kappa_Z^f$, $\rho_Z^f$\footnote{The superscript $f$ denotes the respective fermion.} and $\Delta r$, i.e.

\begin{eqnarray}
\label{Eqn:CorrectedObs}
\MW^2			&=& 	\frac{\MZ^2}{2} (1 + \sqrt{1 - \frac{\sqrt{8} \pi \alpha (1 - \Delta r)}{G_F \MZ^2}} ) \\
\sinfeff		 	&=& 	\kappa_Z^f \sintheta, 		\\
g_V^f 			&=&	 \sqrt{\rho_Z^f} (I_3^f - 2 Q^f \sinfeff)	\\
g_A^f 			&=&	 \sqrt{\rho_Z^f} I_3^f 
\end{eqnarray}

Within the Standard Model, these form factors exhibit a logarithmic dependence on $\MH$, a dependence on quark masses~\footnote{While logarithmic dependencies are important for small quark masses, the quadratic dependence is dominating at large $m_f$, hence implying an overall dominance of the top quark mass contribution.}, dominated by a quadratic dependence of the heaviest quark mass $m_{t}$, and an approximately linear dependence on \MZ, $\alphaQED$ and $\alpha_s$.
Hence, precise measurements of all observables of the EW sector plus the top quark mass, $m_{t}$, and $\alpha_s$, allows a  test of the consistency of the Standard Model or, alternatively, allows a precise prediction of one observable, when all others are known.
This idea of the \textit{global EW fit} has a long history in particle physics and was performed by several groups in the past, e.g. \cite{Bardin:1997xq, Flacher:2008zq, Ciuchini:2014dea, Bardin:1999yd, Akhundov:2013ons, Zyla:2020zbs}. 

The running of the electromagnetic coupling, $\alphaQED$, depends crucially on the loop leptonic and hadronic contributions. 
However, the leptonic and top-quark vacuum polarisation contributions are precisely known or negligible and only the hadronic contribution for the five lighter quarks, \DaHadZfive adds significant uncertainties. 
Hence the electromagnetic coupling $\alphaQED$ is typically replaced by \DaHadZfive within the EW fit.

In the following, we use the Gfitter framework \cite{Haller:2018nnx, Baak:2012kk} to evaluate the impact of the \DaHadZfive observable in the context of the overall fit. 
In particular, we indirectly determine \DaHadZfive using state of the art measurements of the relevant EW precision observables, but also test its impact on the prediction of other observables such as the W boson mass, $M_W$, the Higgs boson mass, $M_H$, and the effective EW mixing angle, $\sin^2(\theta_{eff})$. 
The Gfitter framework includes for the predictions of $M_W$ and $\sin^2(\theta_{eff})$ the complete two-loop corrections and allows for a rigorous statistical treatment. 
For example, it is possible to introduce dependencies among parameters, which can be used to parametrise correlations due to common systematic errors, or to rescale parameter values and errors with newly available results. 
This is relevant for the study of \DaHadZfive, as it depends on $\aSmZ$. 
The rescaling mechanism of Gfitter allows to automatically account for arbitrary functional interdependencies between an arbitrary number of parameters \cite{Flacher:2008zq}.

\section{Including the correlations between \amu and \alphaQED in the EW fit}
\label{Sec:ApproachesCorrelations}

\begin{table*}[tb]
\centering
\small
\begin{tabular}{l |c|c|c}
\hline
\hline
Computation (Energy range)   & \amuHVPLO $[10^{-10}]$& \DaHadZ $[10^{-4}]$ & $\rho$ \\
\hline
Phenomenology (Full HVP) &  $ 694.0 \pm 4.0 $  &  $ 275.3 \pm 1.0 $  & $44\%$  \\
\hline
Phenomenology ($[{\rm Th.};1.8\gev]$)&  $ 635.5 \pm 3.9 $  &  $ 55.4 \pm 0.4 $ & $86\%$ \\
\hline
Phenomenology ($[{\rm Th.};1\gev]$)&  $ 539.8 \pm 3.8 $  &  $ 36.3 \pm 0.3 $ & $99.5\%$ \\
 \hline
Lattice  (Full HVP)      &  $ 712.4 \pm 4.5 $  &  -  &  -  \\
\hline
\hline
\end{tabular}
\caption{ Values of the \amuHVPLO and \DaHadZ integrals computed in either the full energy range~(``Full HVP") or some restricted region, through either a phenomenological approach using experimental hadronic spectra~\cite{Davier:2019can} or with Lattice QCD~\cite{Borsanyi:2020mff}. Where relevant, $\rho$ indicates the correlation coefficient of the uncertainties of the two phenomenological dispersion integrals.}
\label{Tab:Inputs}
\end{table*}

In order to study the impact of the recent \amu-related results on the EW fit, we consider three different approaches.
They all involve correlated shifts of the \linebreak\amuHVPLO and \DaHadZ values, while taking into account the fact that the kernels involved in the dispersion integrals emphasise lower~(higher) energy regions of the hadronic spectra for \amuHVPLO~(\DaHadZ).
However, the methodology and the underlying assumptions are different for each of the three approaches, which is important in the current context where the source of the tension between the various \amu results is unknown.

The values of the HVP contribution integrals used in this study, computed either through a phenomenological approach or through Lattice QCD, are summarised in Table~\ref{Tab:Inputs}, for either the full HVP contribution or more restricted energy ranges.
The latter are starting from the energy threshold~(Th.) and go up to either $1 {\rm~or~} 1.8\gev$, a region where the sum of 32 exclusive hadronic production channels is used for the phenomenological calculation~\cite{Davier:2019can}.
It is to be noted that for the \amuHVPLO dispersion integrals in the phenomenological approach the low energy part dominates for both the central value of the integral and its uncertainty, while for \DaHadZ the high energy regions bring larger contributions.
These are direct consequences of the different energy dependencies for the corresponding integration kernels.
The correlation coefficient $\rho$ between the uncertainties of the two dispersion integrals~(due to the use of the same input hadronic spectra, pQCD and narrow resonance contributions, with different integration kernels) are also indicated, for the various energy ranges that are considered here.
It amounts to $44\%$ when computing the dispersion integrals for the full energy range~\cite{Davier:2019can} and is further enhanced when considering the contributions from lower mass ranges only.
We also note that the full \amuHVPLOpheno contribution obtained in Ref.~\cite{Aoyama:2020ynm} through the conservative merging of several results~$( 693.1 \pm 4.0) \cdot10^{-10}$ is similar to the corresponding value from Table~\ref{Tab:Inputs}, in terms of both the central value and uncertainty.
In addition, in this study we use the difference between \amuexp and \amuSMpheno amounting to $26.0\cdot10^{-10}$~\cite{Davier:2019can}, impacted in particular by the statistical~(systematic) experimental uncertainties of the \amu measurement of $5.4~(3.3)$, in the same units of $10^{-10}$.

In the \approach[0] we apply the same scaling factor for the contributions~(from some energy range of the hadronic spectrum) to the \amuHVPLO and \DaHadZ phenomenological values determined from dispersion integrals, in order to reach some ``target" value for \amuHVPLO.
This scaling can be modelled e.g. as a change of normalisation of the inclusive hadronic spectrum in the corresponding energy range, which is in this sense similar to the studies done in Refs.~\cite{Passera:2008jk,Crivellin:2020zul,Keshavarzi:2020bfy}.
The EW fit is then performed using as input the shifted \DaHadZpr value, with the corresponding uncertainty\footnote{We do not apply here a relative scaling of the \DaHadZ uncertainties for \approach[0], because in case such corrections would be necessary for the central values it is not obvious that the uncertainties would be expected to scale accordingly. Furthermore, even in cases when the scaling is applied only to (part of) the range covered by exclusive channels and the scale factor is therefore at the few percent level, the relative impact on the total \DaHadZ uncertainty would be small.}.

For the \approach[1] the goal is to include \amuHVPLO in the EW fit, using the information on the correlations between the uncertainties of \amuHVPLO and \DaHadZ.
The covariance matrix of the two quantities can be described by a set of two uncertainty components, often called ``nuisance parameters"~(NPs), each of them being fully correlated between the two quantities, but the two being independent between each-other.
There is indeed an infinite number of ways of performing such description of the information in the covariance matrix using two NPs.
One set of such NPs that is especially interesting in this case has the format indicated in Table~\ref{Tab:NPs}, the key point being that \NP[1] impacts both quantities, while \NP[2] only impacts \DaHadZ.
One can evaluate the number of standard deviations by which \NP[1] has to be shifted, in order for the \amuHVPLO determined from dispersion integrals to reach some ``target" value.
The same relative shift of the \NP[1] can then be applied to \DaHadZ.
This shifted value \DaHadZpr is used as input for Gfitter, with the uncertainty provided by the \NP[2]~(which impacts \DaHadZ, but not \amuHVPLO).

\begin{table*}[tb]
\centering
\small
\begin{tabular}{l|c|c}
\hline
\hline
Uncertainty components \hspace{0.5 cm} & \hspace{0.6 cm} \amuHVPLO \hspace{0.6 cm} & \DaHadZ \\
\hline
\NP[1] &  $\sigma(\amuHVPLO)$  &  $\sigma(\DaHadZ) \cdot \rho $  \\
\hline
\NP[2] &  $0$  &  $\sigma(\DaHadZ) \cdot \sqrt{ 1 - \rho^{2} }$  \\
\hline
\hline
\end{tabular}
\caption{ NPs used to describe the covariance matrix of the uncertainties of \amuHVPLO and \DaHadZ~(see text). The $\sigma$ in front of various quantities indicates the corresponding uncertainty and $\rho$ their correlation coefficient.}
\label{Tab:NPs}
\end{table*}

While the full \amuSM prediction, which is directly comparable with \amuexp, also involves contributions from higher order hadronic loops, hadronic light-by-light scattering, QED and EW effects, for \amuHVPLO a direct comparison between the phenomenological and Lattice QCD approaches is possible.
Without loss of generality, in the current application of the two approaches above, the ``target" values of the {\it contribution scaling} or {\it uncertainty shift} are chosen to bring the \amuHVPLOpheno contribution derived phenomenologically to the Lattice QCD value \amuHVPLOlatt, or to bring the \amuSM value to the \amuexp, or yet to reach these values minus one standard deviation of the corresponding uncertainties\footnote{For the studies where the ``target" value is \amuexp minus the corresponding uncertainty, we did not include in the definition of this ``target" other uncertainty components~(e.g. from the light-by-light contribution) involved in the $\amuexp - \amuSM$ comparison. Even if (shifts of the values of) such contributions would certainly impact the picture in the $\amuexp - \amuSM$ comparison, exploring the consequences for the EW fit would be too speculative at this stage and remains beyond the goal of our study. We note however that contributions like hadronic light-by-light, impacting \amuSM but not \DaHadZ, reduce the correlations between the two and hence the impact of \amu on the EW fit.}.
Given the different contributions entering in the various energy ranges involved in the \amuHVPLOpheno calculation~\cite{Davier:2019can}, it is difficult to identify a possible effect that would cause a constant global scaling for all of them, although this cannot be fully excluded either.
In view also of the indications from Ref.~\cite{LaurentSeminar2020} and from the updated version of Ref.~\cite{Borsanyi:2020mff}, it is indeed important to perform the studies of the impact on the EW fit for changes of the hadronic spectra in more restricted energy ranges too.
Studies are performed considering scenarios where the {\it contribution scaling} or {\it uncertainty shift} is done either for the full HVP dispersion integral, or for the sum of the exclusive channels from the energy threshold up to 1.8\gev, or yet for their contribution up to 1\gev~\footnote{For the study involving the range between the energy threshold and 1\gev only the \approach[0] is used, because of the existing correlations between the data uncertainties in this range and in the [1 ; 1.8]\gev interval respectively. Indeed, treating (uncertainties from)~the low energy range in \approach[1] independently of the [1 ; 1.8]\gev interval would not be justified, while the coherent treatment of the two intervals would effectively require applying the \approach[1] for the full range up to 1.8\gev. Note also that the relative uncertainties are also rather similar for the \amuHVPLO and \DaHadZ integrals up to 1\gev, while being also strongly correlated. Due to this, the \approach[1] restricted to the range up to 1\gev would anyway yield rather similar results to the \approach[0].}.
These various choices are summarised in Table~\ref{Tab:shiftsAppr0Appr1}.

\begin{table*}[t]
\centering
\small
\begin{tabular}{ c*{5}{|c} }
\hline
\hline
\amuHVPLO shift &  \multicolumn{2}{c|}{\approach[0]}    &  \multicolumn{3}{c}{\approach[1]}   \\
(Energy range) &  Scaling factor  &  \DaHadZpr         &  Shift \NP[1]  &  $\sigma^{\prime}\left(\DaHadZ\right)$  &  \DaHadZpr \\
\hline
\hline
$\amuHVPLOlatt - \amuHVPLOpheno$    &  1.027  &  0.02826 &  4.6  & $9.0 \cdot 10^{-5}$  &  0.02774 \\
(Full HVP)                     &    &    &    &    &    \\
\hline
$(\amuHVPLOlatt-1\sigma) - \amuHVPLOpheno$    &  1.020  &  0.02808 &  3.5  &  $9.0 \cdot 10^{-5}$  &  0.02769 \\
(Full HVP)                     &    &    &    &    &    \\
\hline
$\amuHVPLOlatt - \amuHVPLOpheno$    &  1.029  &  0.02769 &  4.7  &  $9.5 \cdot 10^{-5}$  &  0.02768 \\
($[{\rm Th.};1.8\gev]$)                     &    &    &    &    &    \\
\hline
$(\amuHVPLOlatt-1\sigma) - \amuHVPLOpheno$    &  1.022  &  0.02765 &  3.5  &  $9.5 \cdot 10^{-5}$  &  0.02764 \\
($[{\rm Th.};1.8\gev]$)                     &    &    &    &    &    \\
\hline
$\amuHVPLOlatt - \amuHVPLOpheno$    &  1.034  &  0.02765 &  -  &  -  &  -  \\
($[{\rm Th.};1\gev]$)                     &    &    &    &    &    \\
\hline
\hspace{0.6cm}$(\amuHVPLOlatt-1\sigma) - \amuHVPLOpheno$  \hspace{0.6cm}  &  1.026  &  0.02762 &  -  &  -  &  -  \\
($[{\rm Th.};1\gev]$)                     &    &    &    &    &    \\
\hline
\hline
$\amuexp - \amuSMpheno$    &  1.037  &  0.02856 &  6.6  &  $9.0 \cdot 10^{-5}$  &  0.02782 \\
(Full HVP)                     &    &    &    &    &    \\
\hline
$(\amuexp-1\sigma) - \amuSMpheno$    &  1.028  &  0.02831 &  5.0  &  $9.0 \cdot 10^{-5}$  &  0.02775 \\
(Full HVP)                     &    &    &    &    &    \\
\hline
$\amuexp - \amuSMpheno$    &  1.041  &  0.02776 &  6.6  &  $9.5 \cdot 10^{-5}$  &  0.02774 \\
($[{\rm Th.};1.8\gev]$)                     &    &    &    &    &    \\
\hline
$(\amuexp-1\sigma) - \amuSMpheno$    &  1.031  &  0.02770 &  5.0  &  $9.5 \cdot 10^{-5}$ &  0.02769 \\
($[{\rm Th.};1.8\gev]$)                     &    &    &    &    &    \\
\hline
$\amuexp - \amuSMpheno$    &  1.048  &  0.02771 &  -  &  -  &  -  \\
($[{\rm Th.};1\gev]$)                     &    &    &    &    &    \\
\hline
$(\amuexp-1\sigma) - \amuSMpheno$    &  1.036  &  0.02766 &  -  &  -  &  -  \\
($[{\rm Th.};1\gev]$)                     &    &    &    &    &    \\
\hline
\hline
\end{tabular}
\caption{ Scaling factors of the hadronic spectra in \approach[0], shifts applied to \NP[1]~(in terms of a number of standard deviations) and the uncertainty to be used in the EW fit $\sigma^{\prime}\left(\DaHadZ\right)$~(which incorporates \NP[2] and the uncertainty from high mass contributions if a restricted range is used for the uncertainty shift, hence including the \aS-related uncertainty too) in \approach[1], together with the corresponding modified \DaHadZpr values, for various shifts of the \amuHVPLO. The latter are achieved through a {\it contribution scaling}~(\approach[0]) or {\it uncertainty shift}~(\approach[1]), applied for various energy ranges of the hadronic spectrum~(see text). The ``$-1\sigma$" following various quantities indicates a subtraction of one standard deviation of the corresponding \amuHVPLO or \amuexp uncertainty. For \approach[0] the uncertainty indicated for ``Full HVP" in Table~\ref{Tab:Inputs} applies to all the configurations presented here, while distinguishing the \aS-related uncertainty and its complementary part~(see text). }
\label{Tab:shiftsAppr0Appr1}
\end{table*}

%\clearpage

\approach[2] consists in performing the EW fit including \amuHVPLO as an extra free parameter, constrained by both the phenomenological value from dispersion integrals~\amuHVPLOpheno and by the Lattice QCD value~\amuHVPLOlatt.
The correlations between the uncertainties of the \linebreak \amuHVPLOpheno and \DaHadZ dispersion integrals for the full energy range~\cite{Davier:2019can} are taken into account in the fit.
The uncertainty due to the finite precision of \aS~(entering here through the pQCD part of the dispersion integrals in the phenomenological approach) also impacts other quantities in the EW fit and is therefore treated separately.
It amounts to $0.14\cdot10^{-10}$~($0.41\cdot10^{-4}$) for \amuHVPLOpheno~(\DaHadZ) and is treated as fully correlated between the two quantities, as well as with the other \aS-related uncertainties in the fit.
The remaining uncertainties are of $3.96\cdot10^{-10}$ for \amuHVPLOpheno and $0.91\cdot10^{-4}$ for \DaHadZ, with a correlation coefficient of $47\%$ between the two.
In the Lattice QCD calculation, pQCD is used in the range $\Qs > 3\gev^2$ for an \amuHVPLO contribution that amounts to $0.16\cdot10^{-10}$, while its uncertainty due to the finite precision of \aS is negligible~\cite{Borsanyi:2020mff}.

\approach[2] brings a slightly improved treatment of the uncertainties and correlations compared to \linebreak\approach[1], where the total covariance matrix~(including the \aS-related uncertainty) has to be used when computing the \NP[1] that impacts \amuHVPLO. There, the \aS uncertainty impacting \DaHadZ is treated as a sub-component of \NP[2]~(and further correlated with other quantities in the EW fit), which effectively de-correlates it from the corresponding uncertainty of \amuHVPLO. This approximate treatment in \approach[1] is however well justified, given the relatively small contribution of the \aS uncertainty to \amuHVPLO.

It is worth noting that in the various scenarios displayed in Table~\ref{Tab:shiftsAppr0Appr1} the scaling factors applied in \linebreak\approach[0] go well beyond the (sub-percent level) systematic uncertainties of the modern experimental measurements of hadronic spectra, used in the phenomenological dispersion integrals.
For this reason we also do not consider applying \approach[0] in more restricted energy ranges below $1~\gev$, as the resulting scaling factors would be even larger and hence unlikely.
Similarly, the shifts of \NP[1]~(expressed as a number of standard deviations) in \approach[1] are relative large, assuming hence that the Gaussian approximation and the correlation coefficients between the dispersion integrals are still valid in this regime.
In \approach[2] the same effect is reflected into a $\chiS$ contribution from the \amuHVPLO component of the fit at the level of about $9.3$ units~(i.e. $3.1$ standard deviations), originating from the tension between \amuHVPLOpheno and \amuHVPLOlatt.
For all these reasons, the current study should not be seen as an attempt to precisely incorporate the \amu inputs into the EW fit, but rather to explore their potential impact under various hypotheses.
Indeed, the three approaches~(with the various choices listed in Table~\ref{Tab:shiftsAppr0Appr1}) allow to probe different hypotheses concerning the possible source(s) of the difference between the phenomenological prediction on one side and the recent Lattice QCD result or the experimental measurement on the other, while assessing the corresponding impact on the EW fit.
While \approach[0] considers a simple normalisation scaling of the hadronic spectra~(and is in this sense comparable with some of the approaches used in Refs.~\cite{Passera:2008jk,Crivellin:2020zul,Keshavarzi:2020bfy,deRafael:2020uif}), \ensuremath{\rm \it Approaches \, 1} and {\it 2} use the information on the experimental uncertainties, with their phase-space dependence and correlations, to guide the evaluation of coherent shifts of \amuHVPLO and \DaHadZ.
\section{Results of the EW Fit}

\begin{table*}[tb]
\begin{tabular}{llllll}
\hline
\hline
\multicolumn{6}{l}{\bf LEP/LHC/Tevatron}                               \\ 
\hline
$M_{Z}$~[GeV]\hspace{1.0cm} 			&  $91.188\pm0.002$ \hspace{1.0cm}	&	$R_{c}^{0}$ 				&  $0.1721\pm0.003$ 			&	$M_{H}$~[GeV] 			&  $125.09\pm0.15$ 		\\
$\sigma^{0}_{\rm had}$  [nb]  	&  $41.54\pm0.037$ 		&	$R_{b}^{0}$ 				&  $0.21629\pm0.00066$ 		 	& 	$M_{W}$~[GeV] 			&  $80.380\pm0.013$	\\	
$\Gamma_{Z}$~[GeV] 		&  $2.495\pm0.002$ 		&	$A_{c}$ 					&  $0.67\pm0.027$ 				&	$m_{t}$~[GeV] 				&  $172.9\pm0.5$ 		\\
$A_{l}$ (SLD) 				&  $0.1513\pm0.00207$ 	&	$A_{l}$ (LEP) 				&  $0.1465\pm0.0033$ 			&	$\sin^{2}\theta_{\rm eff}^l$ 	&  $0.2314\pm0.00023$   \\ \cline{5-6} 
$A_{\rm FB}^{l}$ 			&  $0.0171\pm0.001$   	&  $m_{c}$~[GeV] & \multicolumn{1}{l}{$1.27^{+0.07}_{-0.11}$ GeV } & \multicolumn{2}{l}{{\bf After HL-LHC}} \\ \cline{5-6} 
$A_{\rm FB}^{c}$ 			&  $0.0707\pm0.0035$ 	&	$m_{b}$~[GeV] 			&  $4.20^{+0.17}_{-0.07}$ GeV 		& \multicolumn{1}{l}{$M_{W}$~[GeV]}  			&  $80.380\pm0.008$  \\
$A_{\rm FB}^{b}$ 			&  $0.0992\pm0.0016$ 	&	$\alpha_{s}(M_{Z})$ 	 		&  $0.1198\pm0.003$ 			& \multicolumn{1}{l}{$\sin^{2}\theta_{\rm eff}^l$}  	&  $0.2314\pm0.00012$    \\
$R_{l}^{0}$ 				& $20.767\pm0.025$ 	&	\DaHadZfive [$10^{-5}$]	      	&  $2760\pm9$ 				& \multicolumn{1}{l}{$m_{t}$~[GeV] 		}		&  $172.9\pm0.3$   		\\             
\hline
\hline
\end{tabular}
\centering
\caption{\label{tab:EWFitOverview}Input parameters of the EW fit, based on \cite{Erler:2019hds} as well as expected future uncertainties after the high-luminosity LHC phase.}
\end{table*}

\begin{table*}[tbh]
\centering
\tiny
\begin{tabular}{c | cc | cc | cc | cc}
\hline
\hline
\amuHVPLO shift										& \multicolumn{2}{c|}{Nominal}		& \multicolumn{2}{c|}{\approach[0]} 	&  \multicolumn{2}{c|}{\approach[1]}	&  \multicolumn{2}{c}{\approach[2]}	\\
(Energy range)										& \DaHadZpr	& $\chisqsmall$	& \DaHadZpr	& $\chisqsmall$	& \DaHadZpr	& $\chisqsmall$	& \DaHadZpr	& $\chisqsmall$	\\
\hline
													& 0.02753	&	18.6/16		&		-	&	-			&	-		&	-			& 0.02753 & 28.1/17		\\
    &   &   (p=0.29) &    & &       &   &    &  (p=0.04)   \\
\hline
$\amuHVPLOlatt - \amuHVPLOpheno$   				& 	-		&	-			& 0.02826	&	27.6/16		& 0.02774 &	20.3/16 &-	& -\\
 (Full HVP)     &   &   & & (p=0.04) &    &   (p=0.21)    &   &   \\
\hline
$\amuHVPLOlatt - \amuHVPLOpheno$     	& 	-		&	-			& 0.02769 &	19.9/16		& 0.02768 &	19.8/16		& -	& -\\
 ($[{\rm Th.};1.8\gev]$)     &   &   & & (p=0.22) &    &   (p=0.23)    &   &   \\
\hline
$\amuHVPLOlatt - \amuHVPLOpheno$   			& 	-		&	-	& 0.02765 &	19.6/16		& -	&	-		& -	&	-\\
 ($[{\rm Th.};1.0\gev]$)     &   &   & & (p=0.24) &    &       &   &   \\
\hline
$\amuexp - \amuSMpheno$   					& 	-		&	-			& 0.02856 &	33.6/16		& 0.02782 &	21.2/16		&-	&	-\\
 (Full HVP)     &   &   & & (p=0.01) &    &   (p=0.17)    &   &   \\
\hline
$\amuexp - \amuSMpheno$   			& 	-		&	-			& 0.02776 &	20.6/16		& 0.02774 &	20.4/16		& -	&	-\\
 ($[{\rm Th.};1.8\gev]$)     &   &   & & (p=0.19) &    &   (p=0.20)    &   &   \\
\hline
$\amuexp - \amuSMpheno$   			& 	-		&	-			& 0.02771 &	20.1/16		& -	&	-		& -	&	-\\
 ($[{\rm Th.};1.0\gev]$)     &   &   & & (p=0.22) &    &       &   &   \\
\hline
\hline
\end{tabular}
\caption{Different input values of used \DaHadZpr (see also Table \ref{Tab:shiftsAppr0Appr1}) in the global EW fit and the resulting minimal $\chiS$ values as well as the corresponding p-values. }
\label{Tab:EWFitChi2}
\end{table*}

The input parameters of the fit are summarized in Table \ref{tab:EWFitOverview}. 
They include in particular the measurements from the LEP and SLC collaborations, i.e. the mass and width of the Z boson, the hadronic pole cross sections as well as the forward-backward asymmetry parameters. 
The W boson mass and the top-quark mass are based on measurements at the Tevatron and the LHC, while the Higgs Boson mass is only measured at the latter. 
In summary, the floating parameters in the global EW fit within the Gfitter program are the coupling parameters \DaHadZfive and $\aSmZ$, the masses $M_Z$, $m_c$, $m_b$, $m_t$ and $M_H$, as well as four theoretical error parameters.
In \approach[2], the \chiS definition is modified to include \amuHVPLO as an extra free parameter, constrained by both \amuHVPLOpheno and \amuHVPLOlatt with the corresponding uncertainties, adding hence one degree of freedom to the fit.

In a first step, we determine the minimal $\chiS$ of the global EW fit, using various values for \DaHadZfive according the different approaches described in Section \ref{Sec:ApproachesCorrelations}.~\footnote{The \DaHadZfive values are obtained based on the values in Table~\ref{Tab:shiftsAppr0Appr1}, after subtracting the contribution of the top quark to the pQCD calculation, which amounts to $-0.72\cdot10^{-4}$ with a negligible uncertainty.} The results are summarized in Table \ref{Tab:EWFitChi2}.
As discussed earlier, in \approach[2] the tension between \amuHVPLOpheno and \amuHVPLOlatt induces a contribution to the $\chiS$ of about $9.3$ units.

\begin{figure*}[tb]
\centering
\resizebox{0.48\textwidth}{!}{\includegraphics{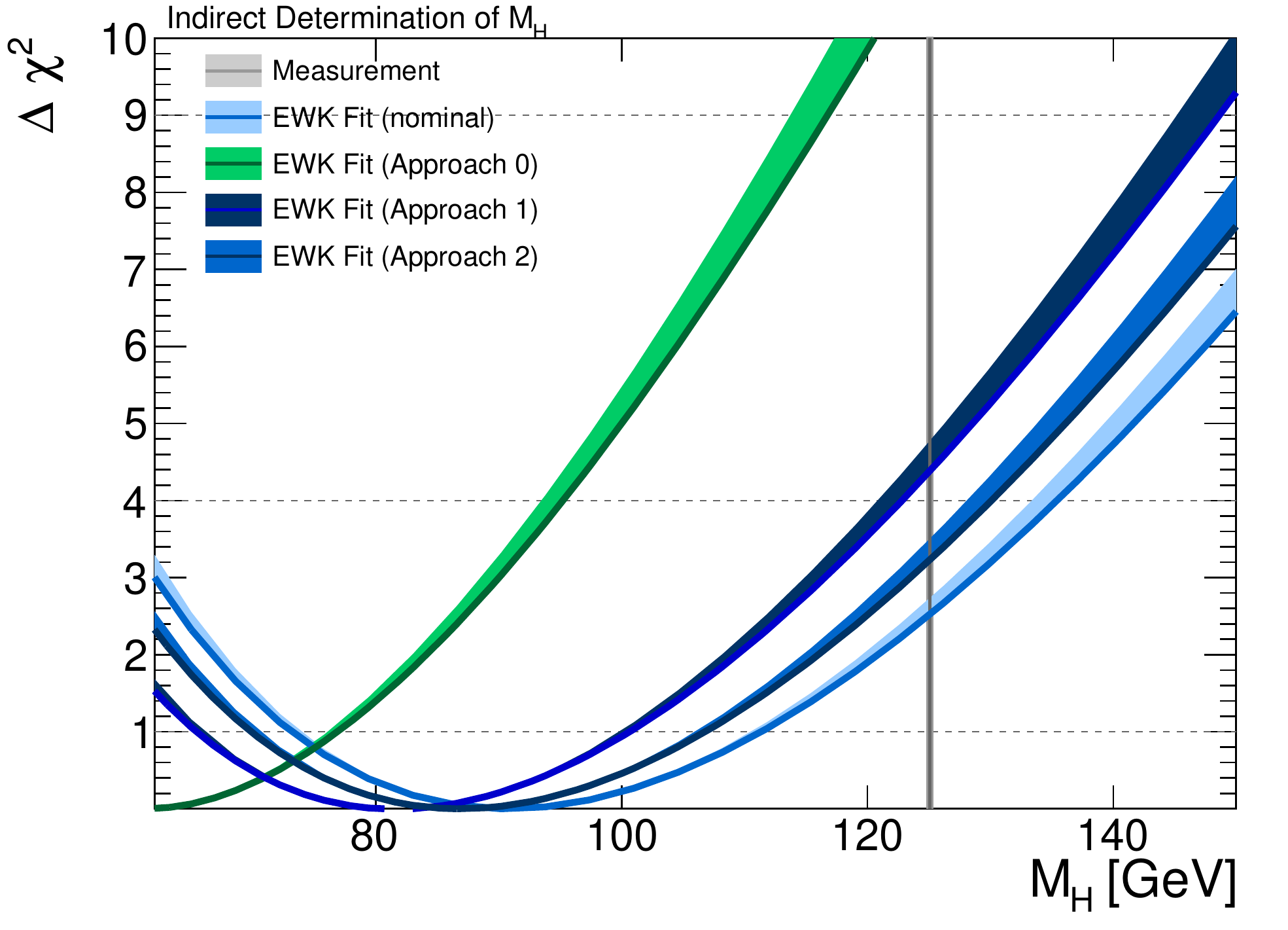}}
\resizebox{0.48\textwidth}{!}{\includegraphics{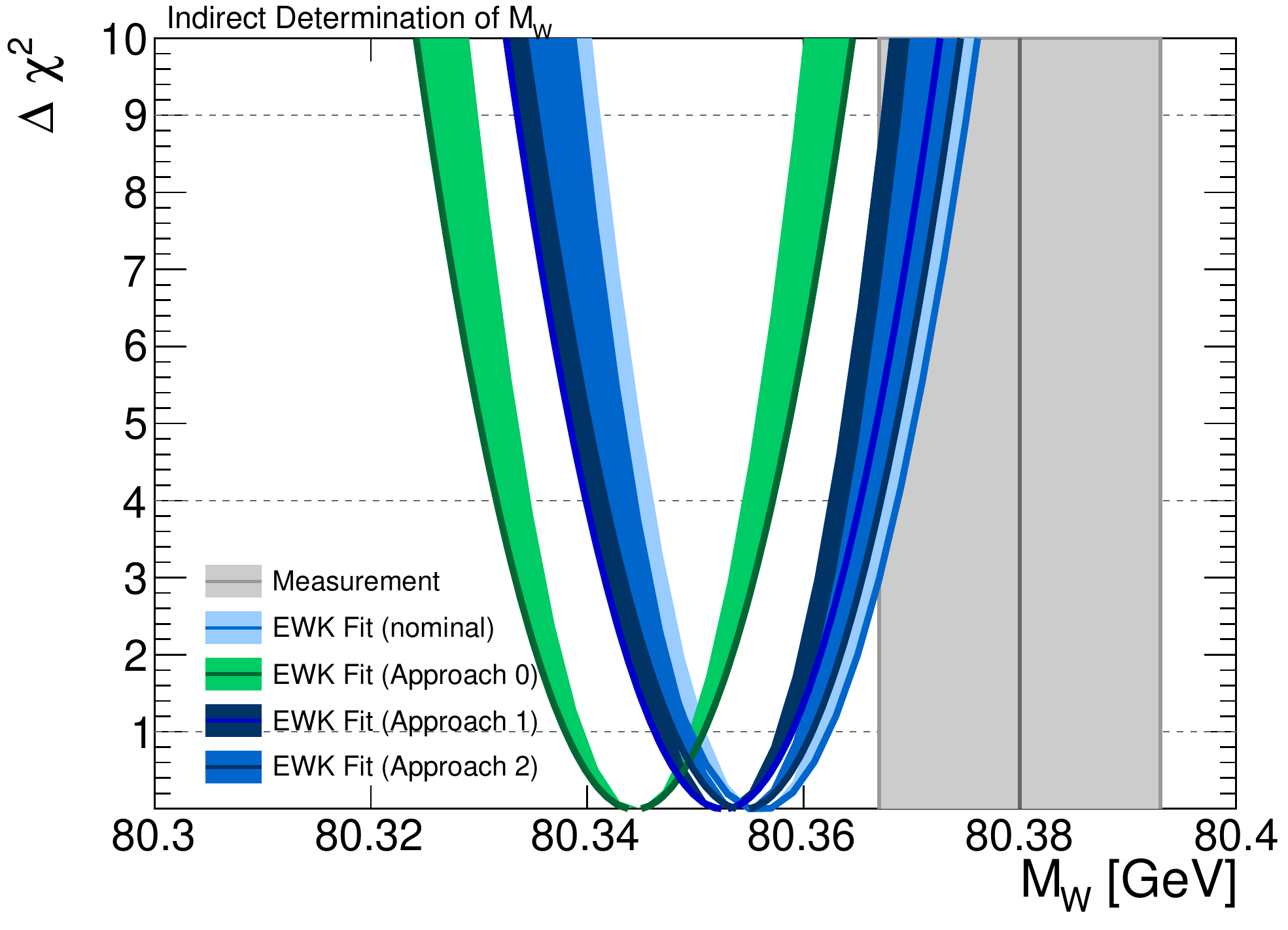}}
\resizebox{0.48\textwidth}{!}{\includegraphics{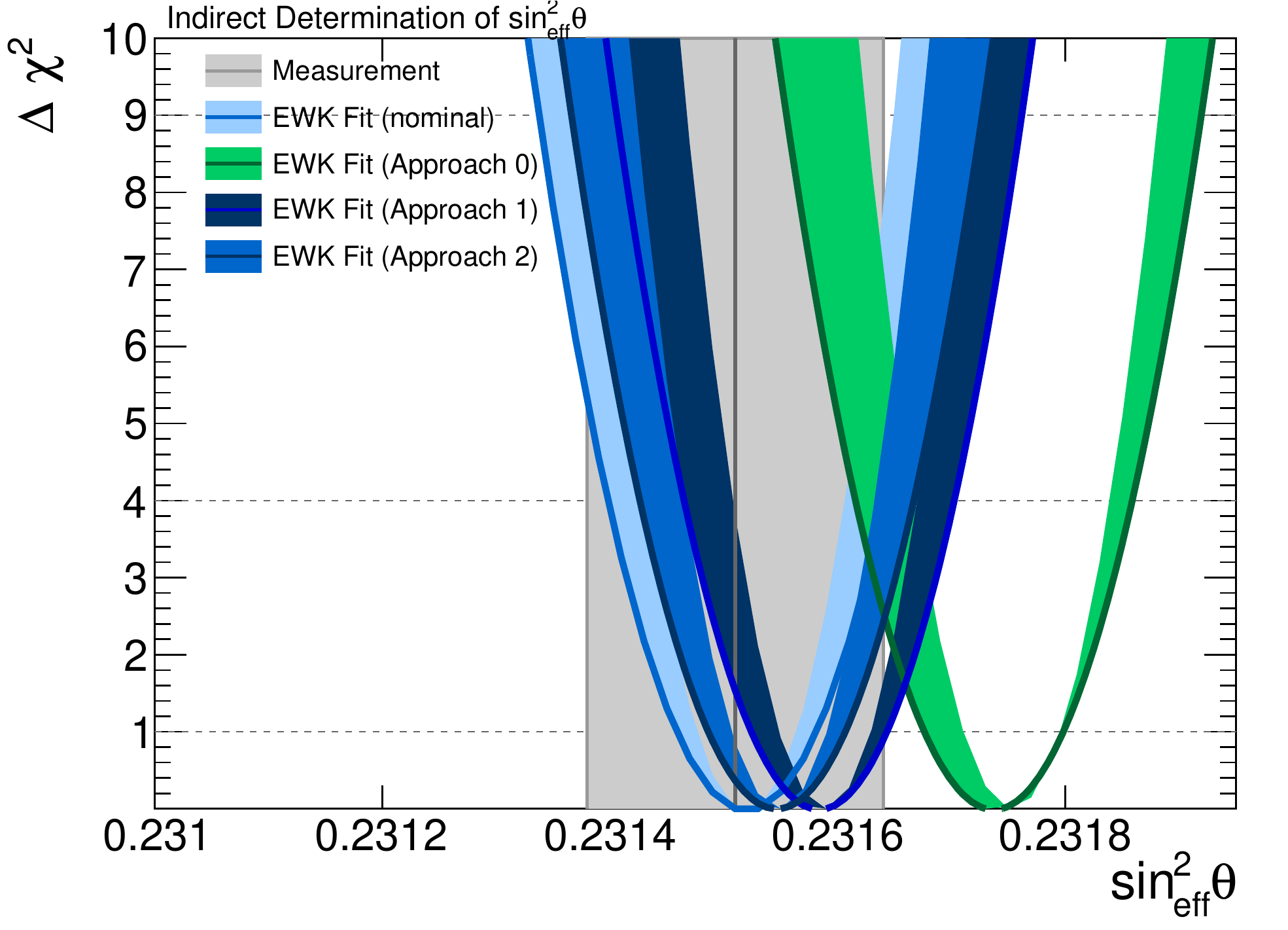}}
\resizebox{0.48\textwidth}{!}{\includegraphics{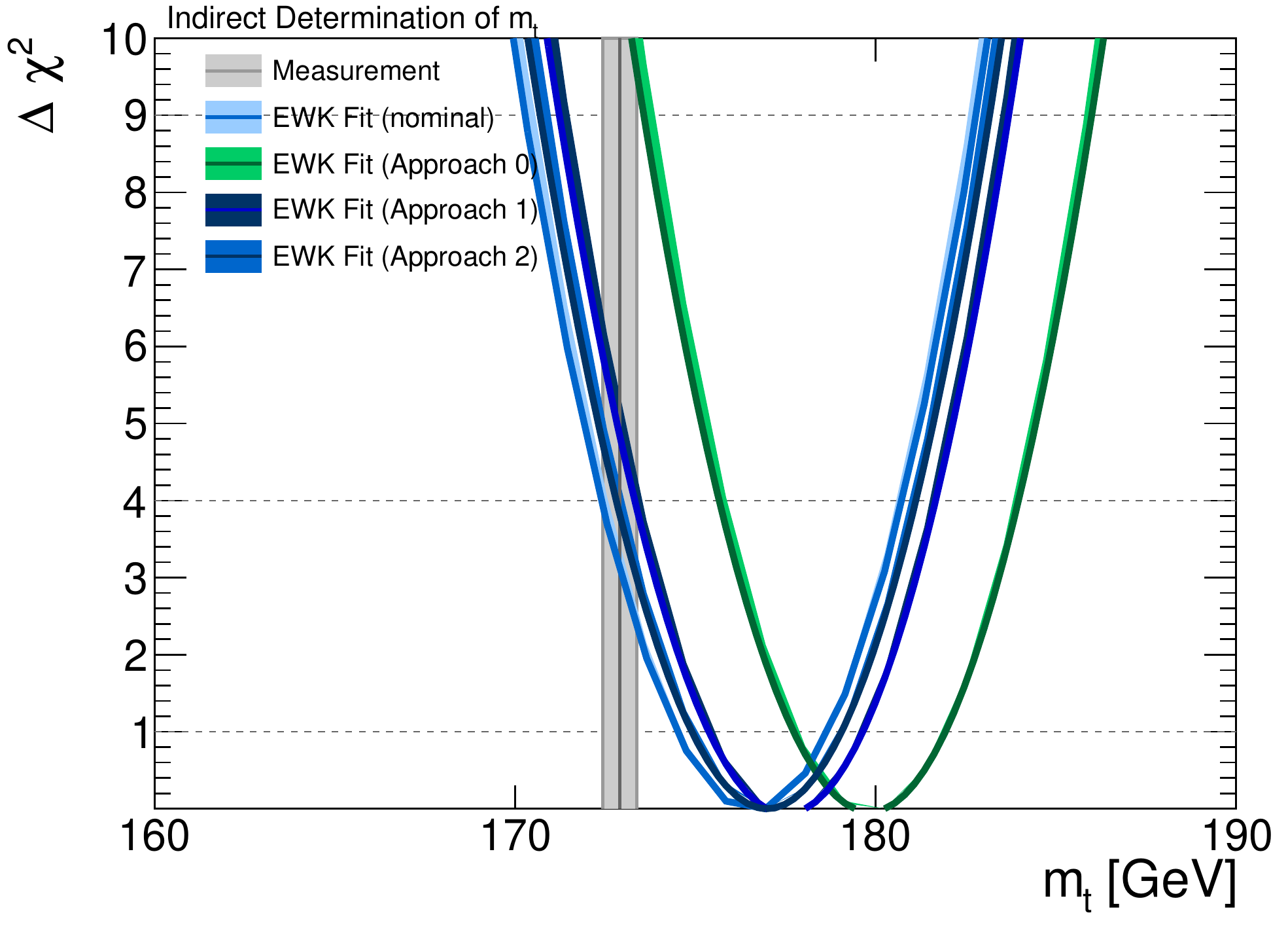}}
\caption{Indirect determination of $M_H$ (upper left), $M_W$ (upper right), $\sin^{2}\theta_{\rm eff}^l$ (lower left) and $m_t$ (lower right) with the global EW fit, using different approaches for \DaHadZ as indicated in Table~\ref{Tab:shiftsAppr0Appr1}, for the $\amuHVPLOlatt - \amuHVPLOpheno$ (Full HVP) case. The shaded bands indicate the theoretical uncertainties within the global EW fit. The measured value and its uncertainty of each observable is indicated as gray vertical band. }
\label{fig:ChiPlot}
\end{figure*}

In a second step, we studied in more detail the $\amuHVPLOlatt - \amuHVPLOpheno$ case and used the three different approaches to indirectly determine several selected observables. Technically, this indirect parameter determination is performed by scanning the parameter in a chosen range and calculating the corresponding $\chiS$ values. 
The value of $\chiS_{\rm min}$ is not relevant for the uncertainty estimation, but only its difference relative to the global minimum, $\Delta \chiS \equiv \chiS - \chiS_{\rm min}$. The $\Delta\chiS=1$ and $\Delta\chiS=4$ profiles define the $1\sigma$ and $2\sigma$ uncertainties, respectively. 
The $\Delta \chiS$ distributions of selected observables ($\MH$, $\MW$, $\sinleff$ and $\mt$) are shown in Figure \ref{fig:ChiPlot}.

When using \approach[0] (comparing either \amuHVPLOlatt and \amuHVPLOpheno, or \amuexp and \amuSMpheno) applying a scaling for the full energy range of the hadronic spectrum the impact on \DaHadZ is large, hence the important shift in the fitted parameters in the EW fit and the corresponding \chiS enhancement.
This is especially significant for \mH, \mW and \mt, where tensions with the measured values are induced in this scenario, which as discussed above is, however, unlikely.
In \ensuremath{\rm \it Approaches \, 1} and {\it 2}, as well as when using \approach[0] with shifts of the HVP contribution applied on more restricted mass ranges, there's less of a change for \DaHadZ and one can conclude that under the corresponding (more realistic) scenarios the impact of the tensions for \amu on the EW fit is small.

\begin{figure*}[tb]
\centering
\resizebox{0.48\textwidth}{!}{\includegraphics{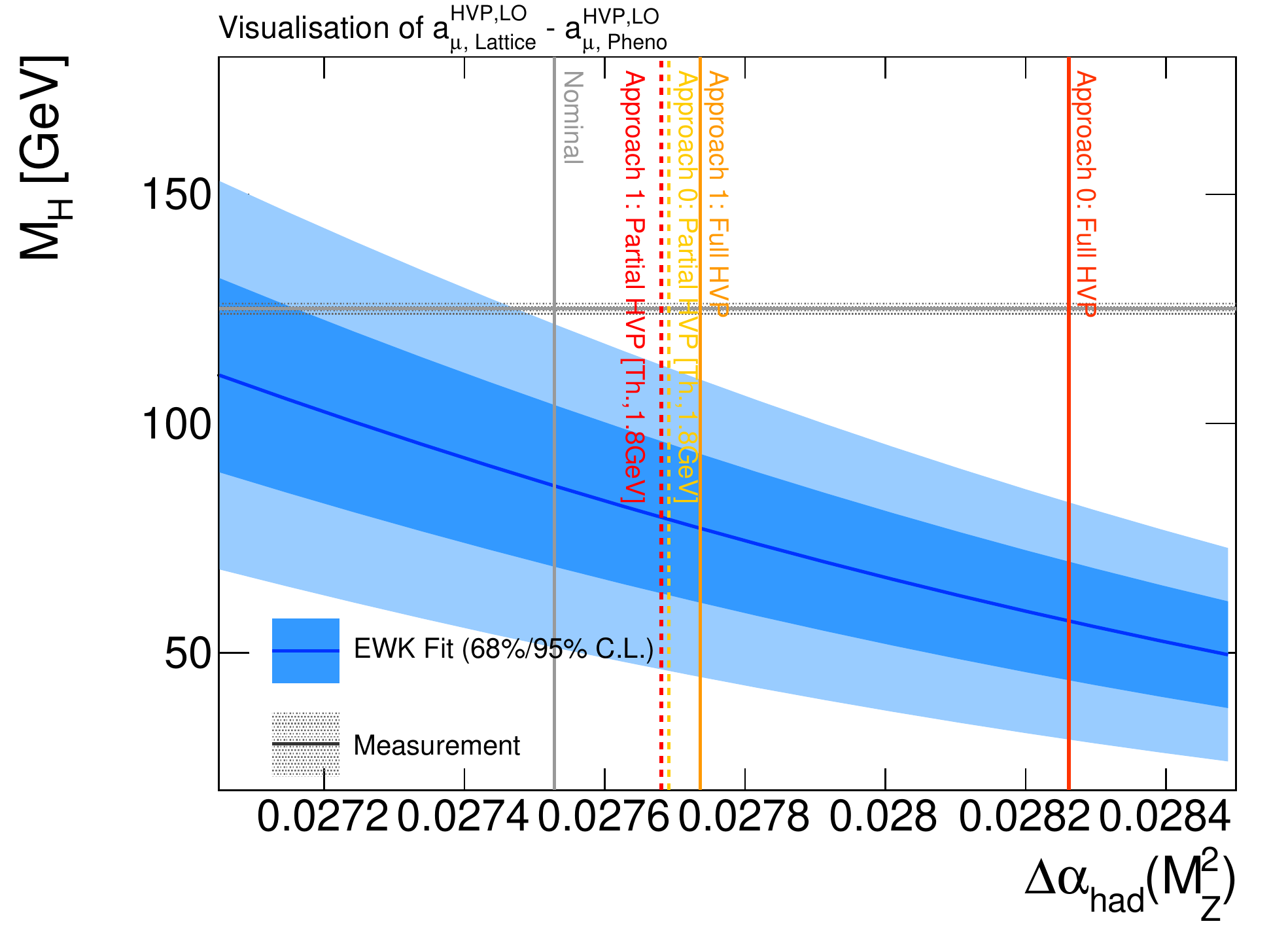}}
\resizebox{0.48\textwidth}{!}{\includegraphics{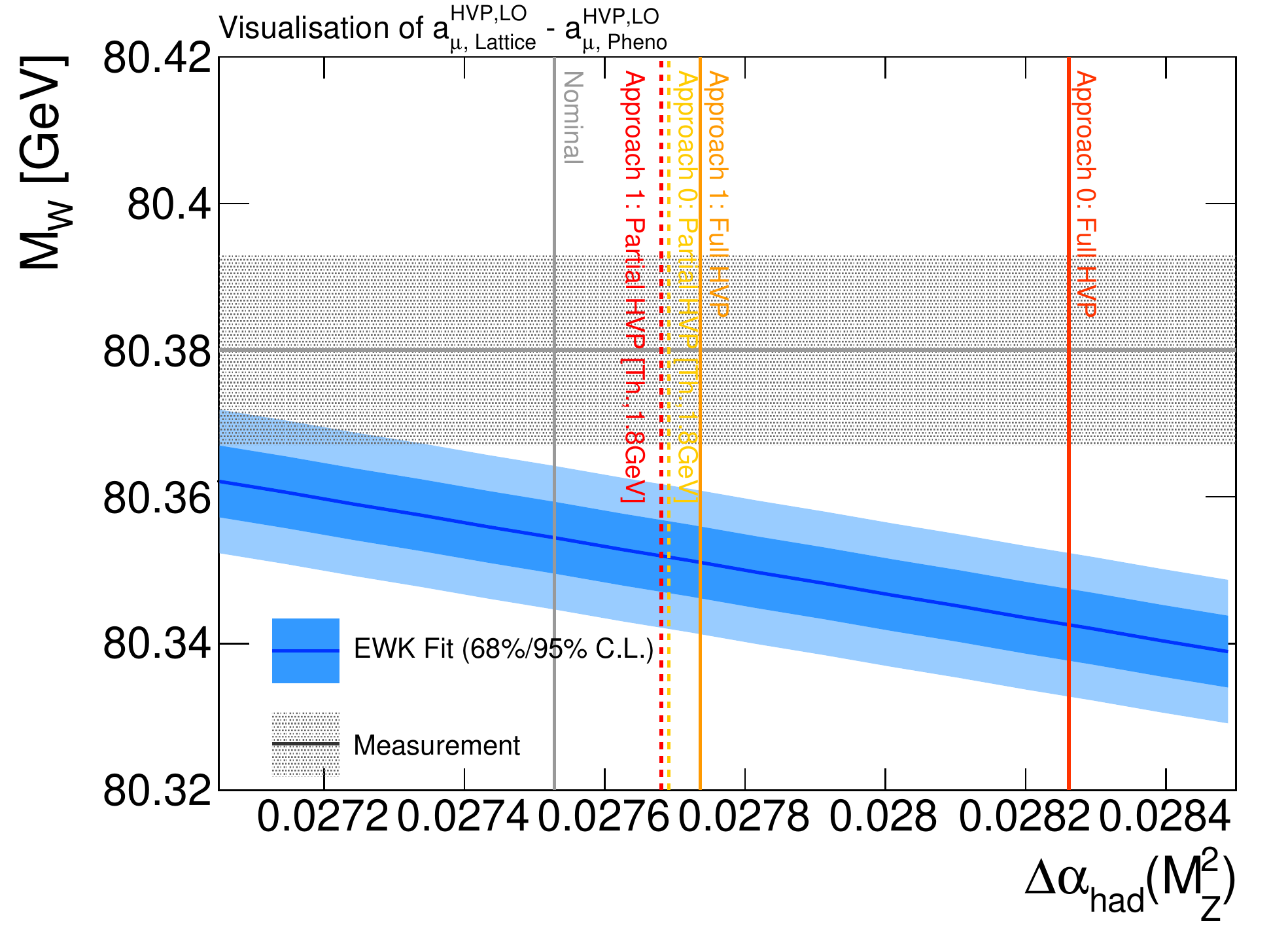}}
\resizebox{0.48\textwidth}{!}{\includegraphics{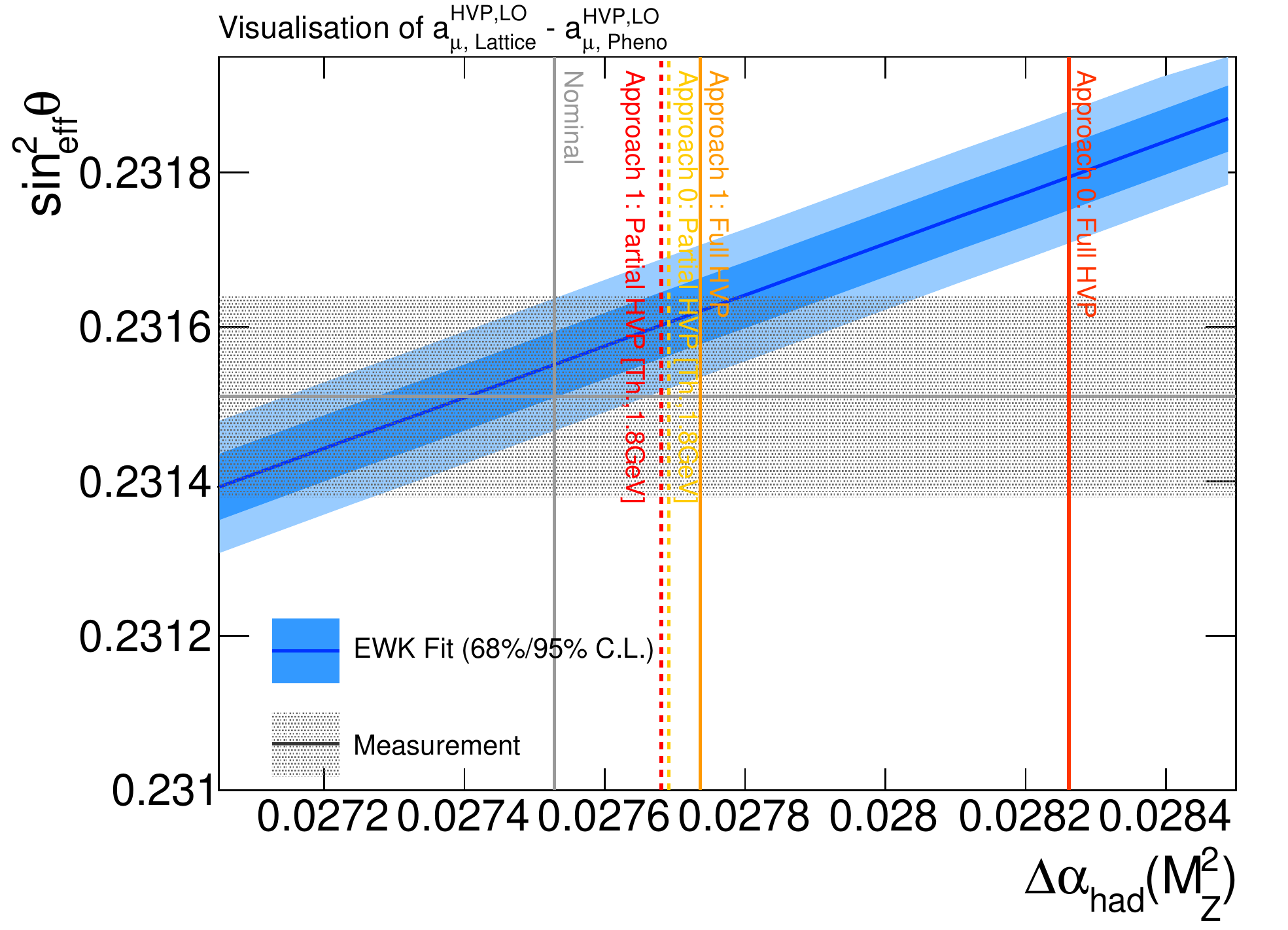}}
\resizebox{0.48\textwidth}{!}{\includegraphics{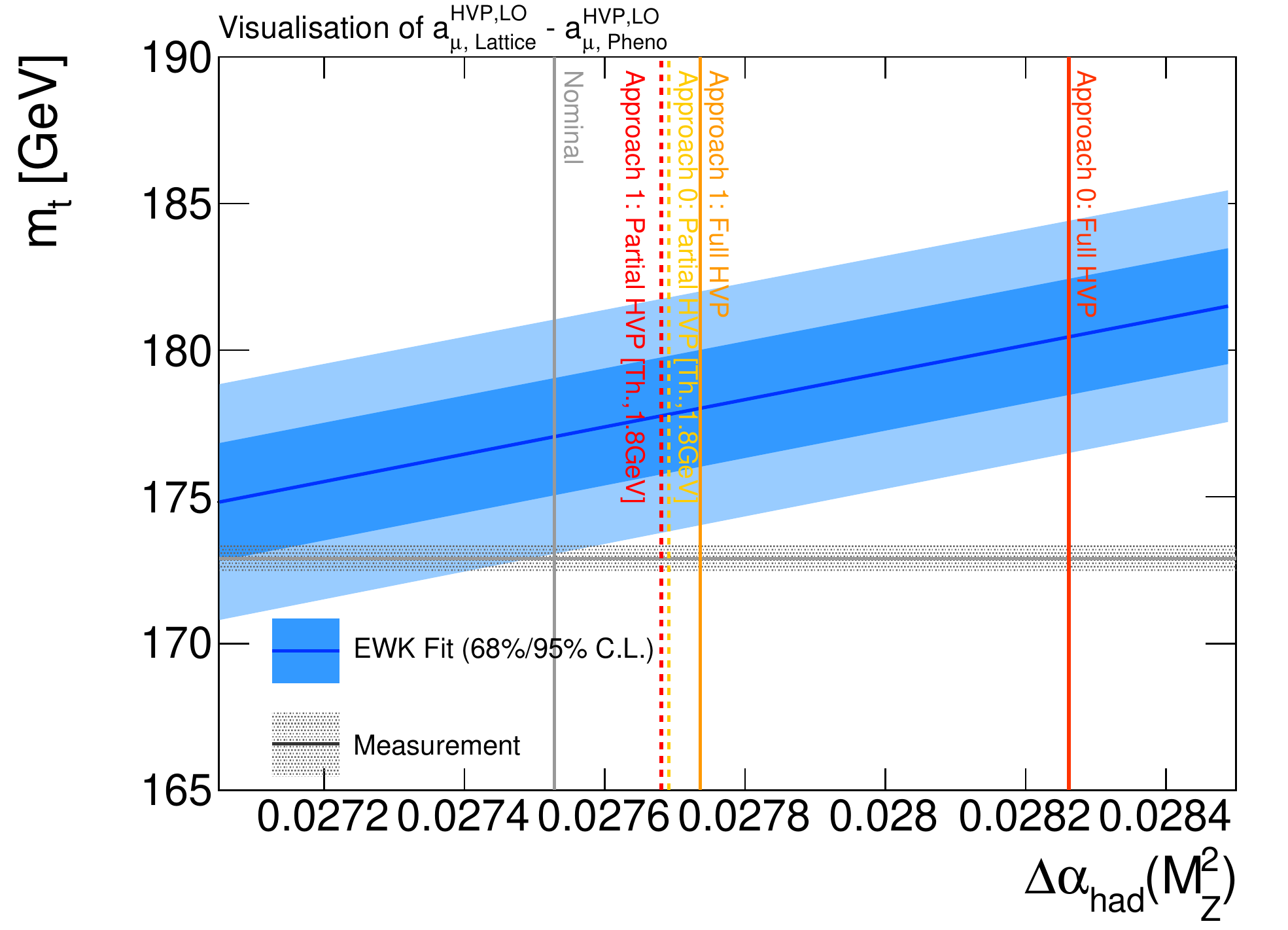}}
\caption{The dependence of the predicted value for $M_H$ (upper left), $M_W$ (upper right), $\sin^{2}\theta_{\rm eff}^l$ (lower left) and $m_t$ (lower right) on \DaHadZ in the global EW fit, together with the corresponding $68\%/95\%$ confidence level (C.L.) intervals, are indicated by the blue bands. The measured value and its uncertainty of each observable is indicated as gray horizontal band. The vertical grey line indicates the result of the nominal fit. The vertical colored lines indicate the results for the \approach[0] and \approach[1], applied either for the full HVP contribution~(continuous lines) or for the range $[{\rm Th.};1.8\gev]$~(dashed lines), for the $\amuHVPLOlatt - \amuHVPLOpheno$ case.}
\label{fig:BandAlpha}
\end{figure*}

The dependence of the predicted value for $M_H$, $M_W$, $\sin^{2}\theta_{\rm eff}^l$ and $m_t$ on \DaHadZ in the global EW fit is illustrated in Figure \ref{fig:BandAlpha}. 
The results of the \approach[0] and \approach[1], applied either for the full HVP contribution or for the range $[{\rm Th.};1.8\gev]$, for the $\amuHVPLOlatt - \amuHVPLOpheno$ case, are also indicated.
The remarks made above about the shifts with respect to the nominal fit result are clearly visible here too.

\begin{figure*}[tb]
\centering
\resizebox{0.48\textwidth}{!}{\includegraphics{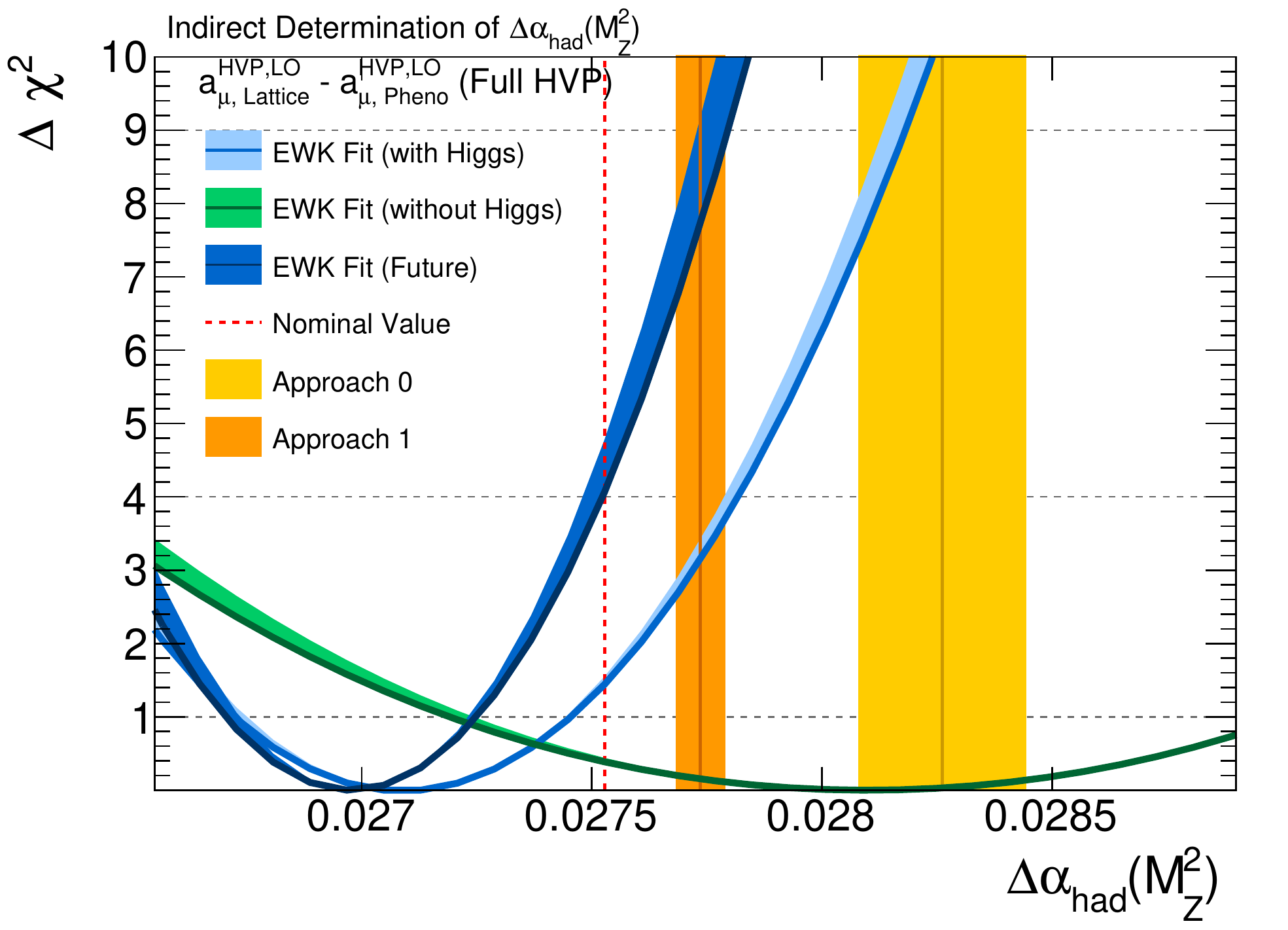}}
\resizebox{0.48\textwidth}{!}{\includegraphics{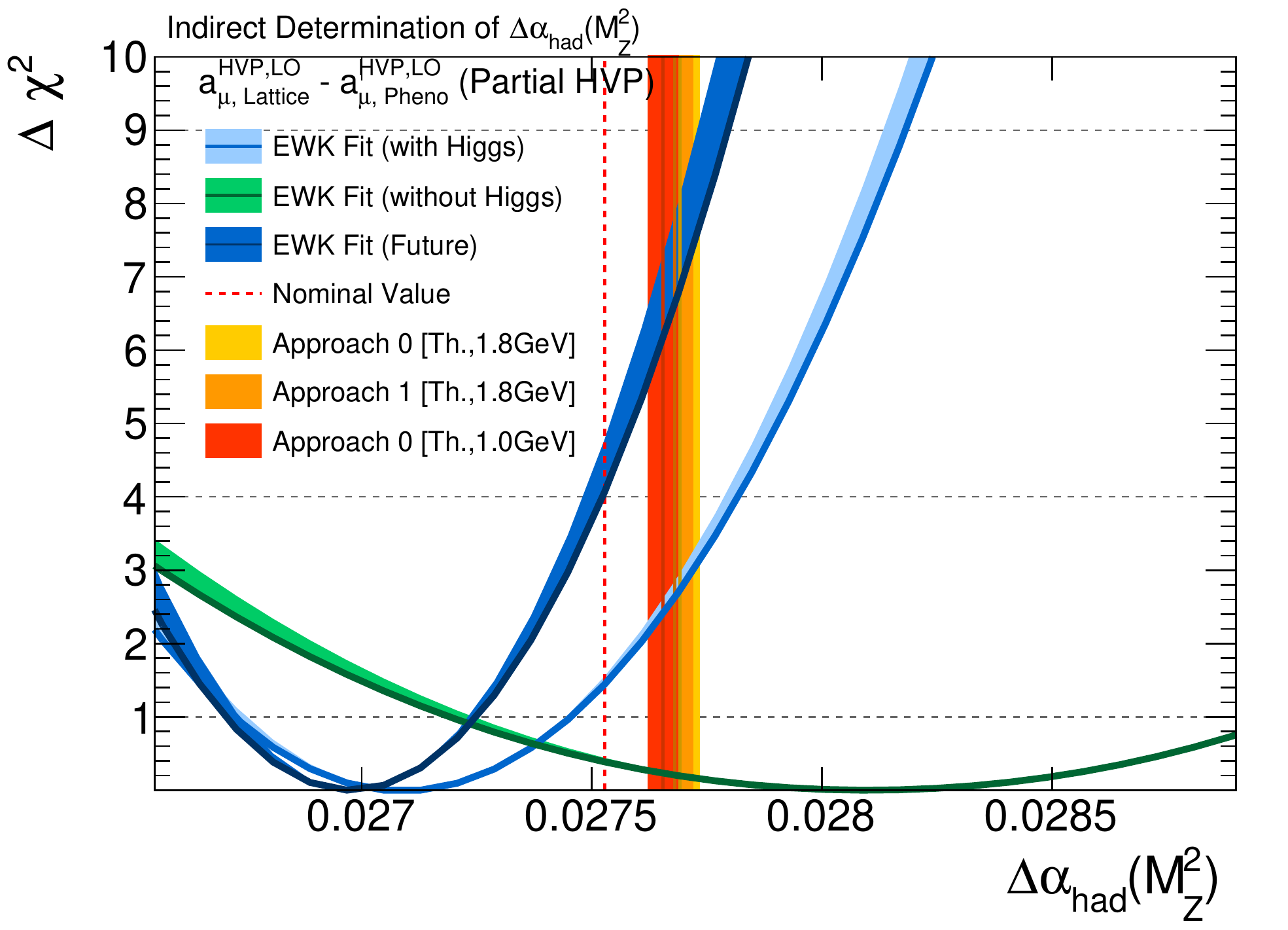}}
\resizebox{0.48\textwidth}{!}{\includegraphics{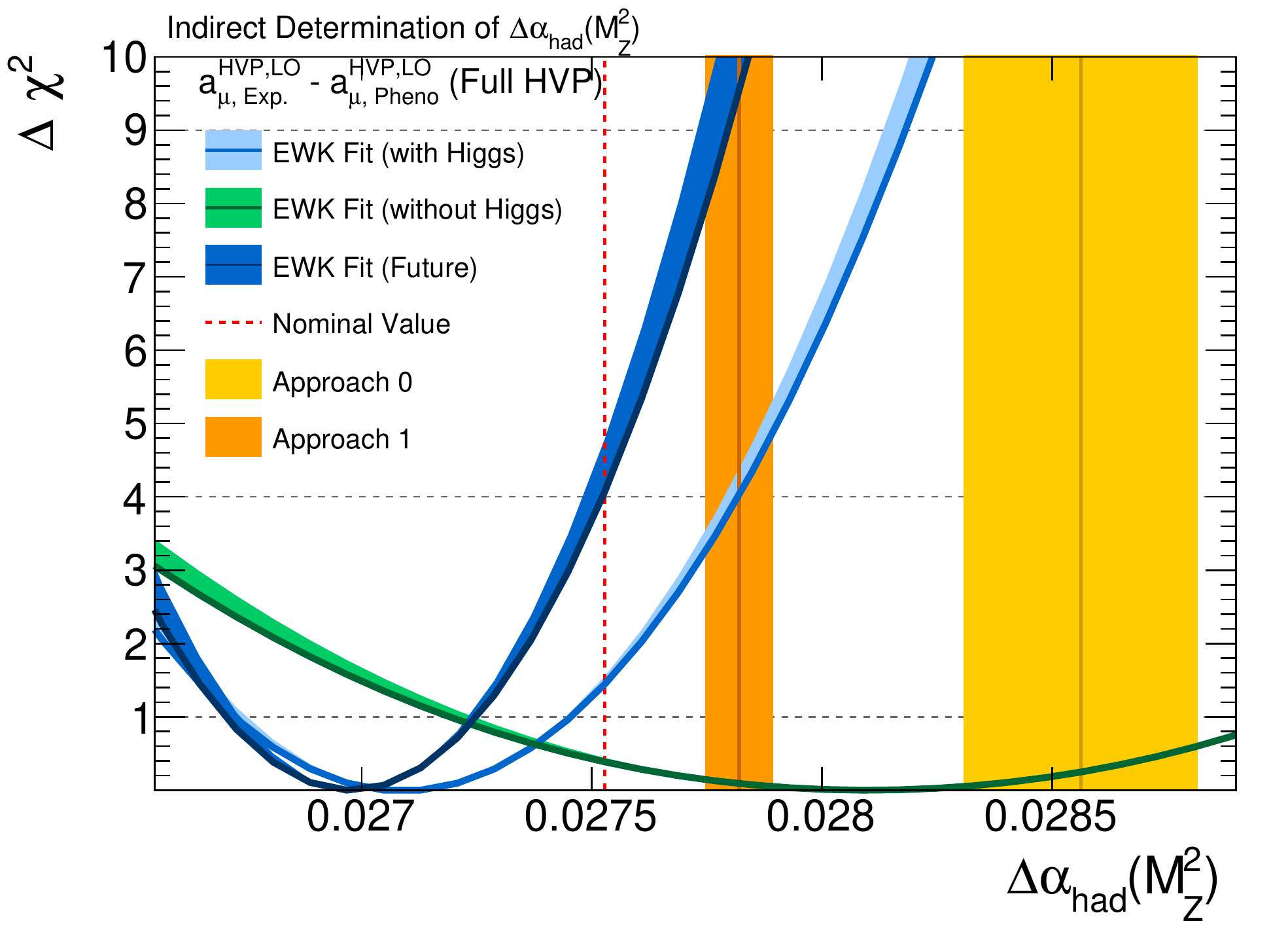}}
\resizebox{0.48\textwidth}{!}{\includegraphics{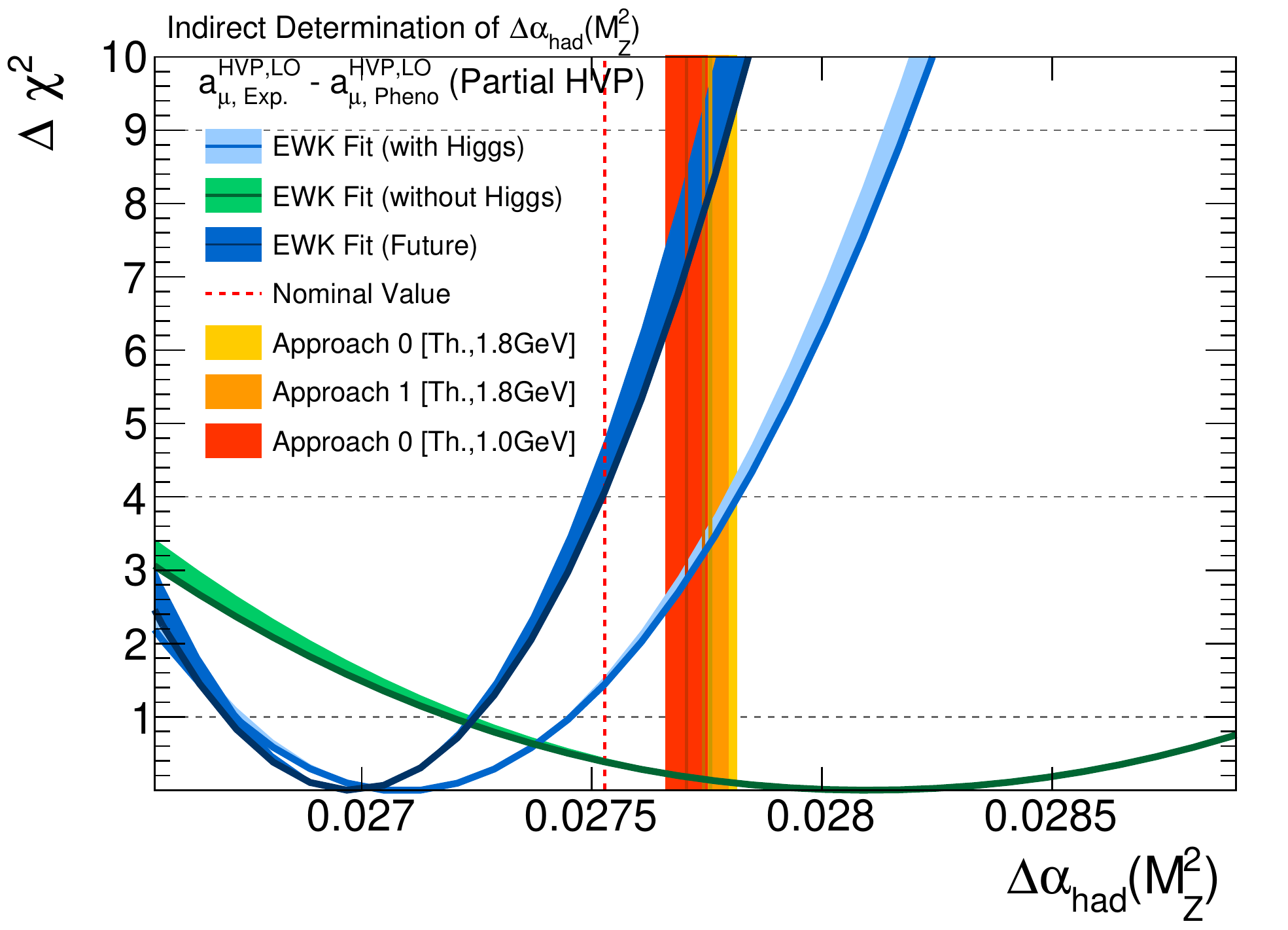}}
\caption{Indirect determination of $\DaHadZ$ and comparison to the different scenarios for $\DaHadZ$ of Table \ref{Tab:shiftsAppr0Appr1}, with the one standard deviation bands indicated therein. The plots correspond to the $\amuHVPLOlatt - \amuHVPLOpheno$ case~(top) and to $\amuexp - \amuSMpheno$~(bottom), for the Full HVP~(left) and for a partial mass range~(right).}
\label{fig:ScanAlpha}
\end{figure*}

Thirdly, we determine indirectly the value of \linebreak\DaHadZfive~(without including any explicit constraint on it in the EW fit) using the other EW observables, including and excluding the Higgs boson mass\footnote{The comparison of the $\chi^2$-distributions when not including the Higgs boson mass in the fit was added to illustrate the impact of the Higgs Boson discovery on the prediction of \DaHadZfive.} as well as assuming improved precisions on the EW observables at the end of the high luminosity LHC phase (see Table \ref{tab:EWFitOverview}). The corresponding $\Delta \chiS$ distributions are shown in Figure \ref{fig:ScanAlpha} for all three cases, yielding values of $\DaHadZfive = 0.02716 \pm 0.00033$ (including $m_{H}$), $\DaHadZfive = 0.02817 \pm 0.00087$ (excluding $m_{H}$) and $\DaHadZfive
 = 0.02706 \pm 0.00025$ (with future measurement precisions), respectively. In addition, we indicate the predicted values of \DaHadZfive, previously discussed in Section \ref{Sec:ApproachesCorrelations}. The uncertainties on these predictions are estimated based on the uncertainty of the ``target" value, driven either by the experimental measurement or the Lattice QCD calculation. 
 
Including the constraint on the Higgs boson mass significantly improves the accuracy of the indirect \linebreak~\DaHadZfive determination.
Then, in all the configurations the tension between the fitted and the predicted \DaHadZ is enhanced, compared to the one for the nominal prediction.
However, here also the tension becomes significant only when using the \approach[0] applying a scaling for the full energy range of the hadronic spectrum.

\section{Summary and Conclusions}

We studied the potential impact on the EW fits of the tensions between the current determinations of the HVP contributions to \amu, based on either phenomenological dispersion integrals of hadronic spectra or respectively on Lattice QCD calculations.
Similarly, we also considered the impact of the current tension between the experimental measurement of \amu and its total theoretical prediction based on the phenomenological calculations of the HVP.
We considered an approach based on coherent shifts of the hadronic spectra in various mass ranges~(comparable with some of the approaches used in Refs.~\cite{Passera:2008jk,Crivellin:2020zul,Keshavarzi:2020bfy,deRafael:2020uif}) and two novel approaches that take into account the correlations between the uncertainties of the theoretical predictions of \amu and of the running of $\alpha_{\rm QED}$ in the phenomenological approach.
It is found that the impact on the EW fit can be large in scenarios involving global shifts of the full HVP contribution.
However, such scenarios seem unrealistic, since they require the same relative shift to be applied in mass ranges of the hadronic contributions where very different methodologies and inputs are being used.
Indeed, recent studies where the \DaHadZ contribution is broken into energy intervals also indicate that the differences between the two evaluations, based on the phenomenological and Lattice QCD calculations respectively, seem to originate from the low energy region~(see Ref.~\cite{LaurentSeminar2020} and the updated version of Ref.~\cite{Borsanyi:2020mff}).
The impact on the EW fit is much smaller if the shift is restricted to a lower mass range~(as also noted in Refs.~\cite{Keshavarzi:2020bfy,deRafael:2020uif}) and/or if the shift of the \amu prediction is propagated to \alphaQED following the pattern of the current uncertainties with their full set of correlations, as done in our two novel approaches.
In the case of the latter scenarios, addressing the current tensions at the level of \amu would not induce significant tensions in the EW fit, implying at most a $2.6/16$ increase in the corresponding $\chisq$, while the changes for the resulting fit parameter values are small too.~\footnote{In the second version of Ref.~\cite{Borsanyi:2020mff} a somewhat smaller difference between the phenomenological and Lattice QCD calculations of \amuHVPLO is observed, reducing even further the potential impact on the EW fit.}

An improved precision for the experimental measurement of \amu, further precise measurements of the hadronic spectra allowing to hopefully also clarify the tension between BABAR and KLOE in the \pp channel, as well as other precise Lattice QCD calculations are expected to become available in the future~\cite{Aoyama:2020ynm}.
Beyond the main goal of exploring the possibility of a contribution from new physics in the comparison between the measurement of \amu and its theoretical predictions, the improved precision will allow to better scrutinise the impact of these findings on the EW fit.

\section*{Acknowledgements}
The authors thank Laurent Lellouch for useful discussions and encouragements during the preparation of this study, as well as for the careful reading of this manuscript.
B.M. acknowledges the fruitful collaboration with our colleagues and friends Michel Davier, Andreas Hoecker and Zhiqing Zhang on the HVP studies.
The authors also thank them for the very interesting feedback provided since the early stages of this project and throughout its preparation. 
M.S. acknowledges in addition the valuable collaboration with the Gfitter group during the past years.

\bibliographystyle{apsrev4-1} %%% physical review (up to date)
\bibliography{./Bibliography.bib}

\end{document}